\documentclass{aa}  

\usepackage{etoolbox}
\AtBeginDocument{\nolinenumbers}
\usepackage{lipsum}

\usepackage{graphicx}
\usepackage{txfonts}
\usepackage{hyperref}
\hypersetup{
    colorlinks=true,
    linkcolor=blue,
    citecolor=blue,
    filecolor=blue,
    urlcolor=blue,
}

\begin{document}
   \title{A refined method for measuring cosmological distances using variability and proper motions in AGN with VLBI-detected counter-jets}

   \author{Jeffrey A. Hodgson\inst{1}\corrauth{jhodgson@sejong.ac.kr}
        \and Anthony Carr\inst{2}
        \and David Parkinson\inst{2,3}
        \and Matteo Statti\inst{4}
        \and Jaehyeon Myeong\inst{1}
        \and Benjamin L'Huillier\inst{1}
        \and Arman Shafieloo\inst{2,3}
        \and Ioannis Liodakis\inst{5,6}
        \and Se-Heon Oh\inst{1}
        }

\institute{Sejong University, 209 Neungdong-ro, Gwangjin-gu, Seoul 05006, Republic of Korea
\and Korea Astronomy and Space Science Institute, 776 Daedeok-daero, Yuseong-gu, Daejeon 34055, Republic of Korea
\and University of Science and Technology, Daejeon 34113, Republic of Korea
\and York University, 4700 Keele Street, Toronto, ON M3J 1P3, Canada
\and Institute of Astrophysics, Foundation for Research and Technology-Hellas, GR-70013 Heraklion, Greece
\and Max-Planck-Institut f\"{u}r Radioastronomie, Auf dem H\"{u}gel 69, D-53121 Bonn, Germany
}

\titlerunning{A refined method for measuring cosmological distances}
\authorrunning{Hodgson et al.}

   \date{Received XX, 20XX; accepted XX, 20XX}
 
  \abstract
   {In a previous paper, we described a `standard speed-gun' (SSG) distance that uses the speed of light to standardize a ruler under the assumption that the radio variability seen in blazars is causally limited. The apparent size is then measured with Very Long Baseline Interferometry in order to derive the angular diameter distance. A key limitation of this method is that it requires knowledge of the relativistic Doppler factor. Previously, we estimated the distance to the bright radio source, 3C 84 at the center of the Perseus cluster assuming a Doppler factor of $\delta \sim$1.}
   {In this paper, we aim to describe how a detected counter-jet and approaching jet proper motions can be used to remove the need for knowledge of the Doppler factor when measuring cosmological distances in this way.} 
   {Under the assumption of a disk (or spherical) geometry and parameterizing the relationship between the physical emitting region and the variability timescale via a causality correction factor ($\kappa$), we estimate a refined angular diameter distance to 3C 84 ($z=0.0178$) with statistical errors.}
   {Assuming $\kappa=1$, we derive distances of $D_{\rm A, \rm disk} =  78.9_{-9.8}^{+11.0}$ Mpc (or  $D_{\rm A, \rm sphere} =  71.2_{-8.8}^{+9.7}$ Mpc). Comparing these results to literature benchmarks, we find that the spherical assumption yields a distance consistent with local Type Ia supernovae calibrated to the SH0ES $H_0$, while a disk-like geometry aligns with expectations from a lower $H_0$ cosmology.}
   {Ultimately, this demonstrates that utilizing jet and counter-jet kinematics successfully removes the Doppler-factor dependence from the standard speed-gun method, providing a viable independent distance estimate once the geometric structure of the jet is resolved.}


   \maketitle

\nolinenumbers

\section{Introduction}
\label{sec:intro}

\begin{figure*}[htbp]
    \centering
    \includegraphics[width=1.0\linewidth]{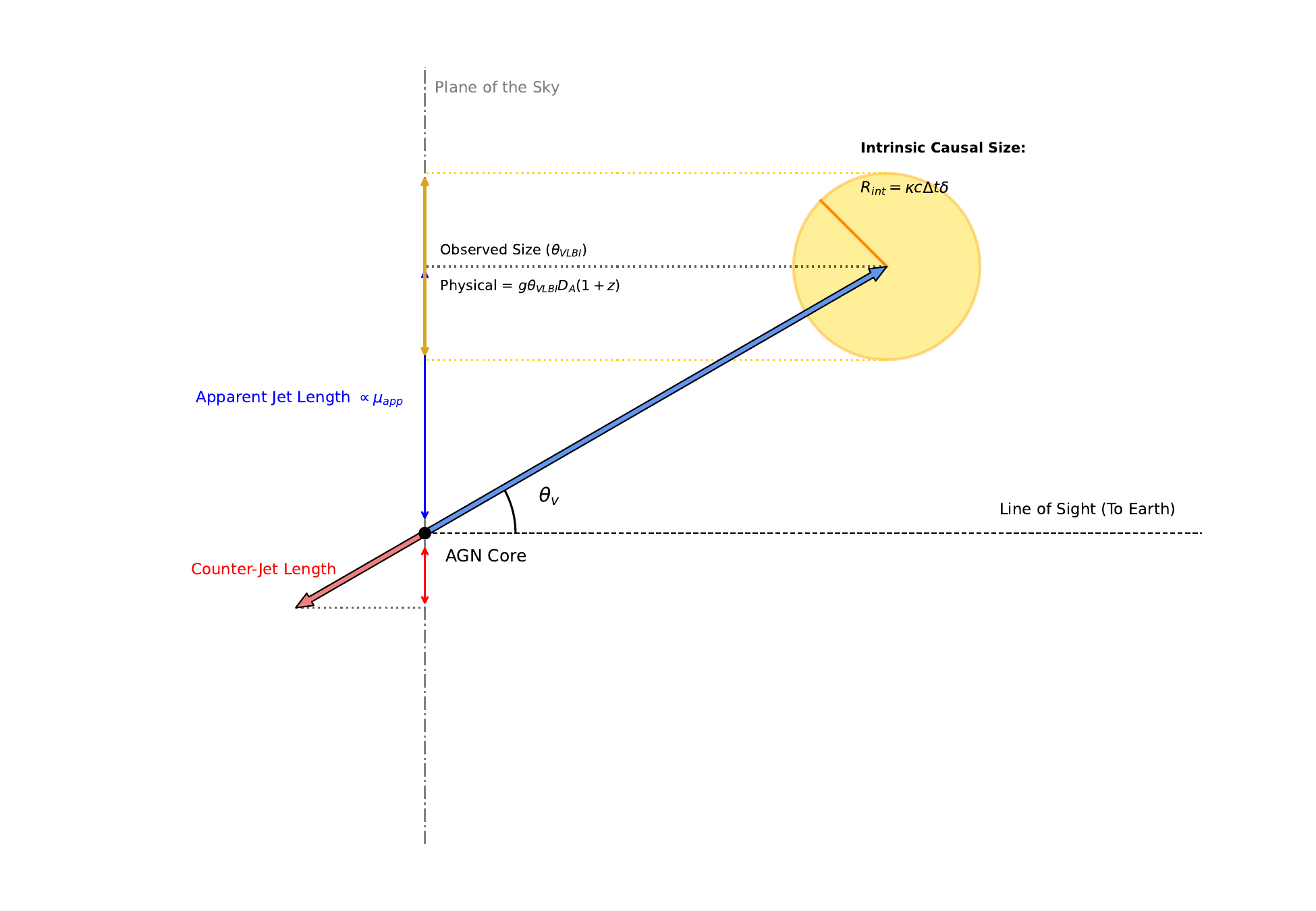}
\caption{\textbf{Schematic representation of the Standard Speed-Gun Counter-Jet (SSG-CJ) method.} The central Active Galactic Nucleus (AGN) core launches identical, bipolar jets at an intrinsic speed $\beta c$ and viewing angle $\theta_{\text{v}}$ relative to the observer's line of sight. \textit{Kinematics (left/y-axis):} Due to light-travel time effects, the approaching jet (blue) appears significantly longer and exhibits a faster proper motion ($\mu_{\text{app}}$) on the plane of the sky compared to the receding counter-jet (red, $\mu_{\text{rec}}$). This asymmetry allows for the measurement of the apparent length ratio, $R_{\text{L}}$. \textit{Causality and Size (right/blob):} The intrinsic physical size of the emitting region (yellow shaded circle) is restricted by the variability timescale to a causal radius of $R_{\text{int}} = \kappa c \Delta t \delta$, where $\kappa$ is the causality correction factor and $\delta$ is the Doppler factor. This physical region projects onto the plane of the sky as an observed angular size, $\theta_{\text{VLBI}}$, which corresponds to a physical size of $g \theta_{\text{VLBI}} D_{\text{A}} (1+z)$, where $g$ is the geometric correction factor. By combining the kinematic length ratio with the causality relation (bottom right panel), the unobservable Doppler factor $\delta$ is analytically eliminated, yielding a direct measurement of the angular diameter distance, $D_{\text{A}}$. }   \label{fig:method_geometry}
\end{figure*}

Accurate calibration of the extragalactic distance scale in the local Universe is critical for resolving the current tension in Hubble Constant ($H_{0}$) measurements. The discrepancy between the value of $H_{0}$ inferred from the Cosmic Microwave Background (CMB) by \citet{planckH0} ($\sim$67.4 km s$^{-1}$ Mpc$^{-1}$) and that measured locally using the Cepheid-Supernova distance ladder by the SH0ES collaboration \citep[$\sim$73.0 km s$^{-1}$ Mpc$^{-1}$;][]{shoesH0} has reached a statistical significance of $>5\sigma$. This tension suggests either unknown systematics in the measurements or new physics beyond the $\Lambda$CDM model. To resolve this, independent distance anchors in the Hubble flow that do not rely on the traditional distance ladder (e.g. Cepheids or the Tip of the Red Giant Branch (TRGB)) are required. Geometric methods, such as water megamasers \citep{pesce2020}, offer a direct measurement, but are rare. Here we further explore a radio-based geometric method for measuring cosmological distances applied to Active Galactic Nuclei (AGN).

It has earlier been proposed to use parsec-scale jets as distance indicators, including using counter-jets \citep{homan2000}.  They showed that by comparing the observed proper motions of jet components with their measured Doppler factors, the angular size distance to a source can be deduced independently of the traditional distance ladder. Building on this, in our previous paper \citet{hodgson2020}, (Paper I), we described the `standard speed gun' (SSG), where the speed of light can be used to calibrate a standard ruler and thus determine the angular diameter distance ($D_{\rm A}$) by measuring the apparent size using Very Long Baseline Interferometry (VLBI). 

Physically, the SSG method relies on causality arguments: if a source changes its brightness over time, the emitting region cannot be physically larger than the distance light travels in that time. In the case of blazars, this is likely to be true, with any deviations from this encapsulated by the causality correction factor $\kappa$. By comparing this intrinsic physical size limit to the observed angular size on the sky, the distance to the source is measured as:
\begin{equation}\label{eq:DA_original}
    D_{\rm A} = \frac{\kappa c\Delta t \delta}{g\theta_{\rm VLBI}(1+z)},
\end{equation}
where $c$ is the speed of light, $\Delta t$ is the variability timescale, $\delta$ is the relativistic Doppler factor, $\theta_{\rm VLBI}$ is the Half-Width-Half-Maximum of a Gaussian fitted to the VLBI image, $g$ is a geometric correction factor to convert a Gaussian to a more realistic geometry (e.g. $\sim$1.6 for a disk or $\sim$1.8 for a sphere) and $z$ is the cosmological redshift. In this work and others, the variability timescale is taken to be the characteristic growth (e-folding) timescale of the rising brightness, and so this method requires high cadence monitoring over months-years.

This is a `single-rung' method of distance measurement, which has the potential to be usable at both low redshift (such as in Paper I) and at high ($z>1.5$) redshift (Myeong et. al. submitted, \citet{song2026}. Given the sensitivity improvements expected for SKA-VLBI, it will be possible to probe much deeper into the universe and detect significantly more sources. For local sources, peculiar velocities can dominate the cosmological expansion, heavily complicating distance measurements. However, as a new single-rung distance indicator that works at both low and high redshift, the SSG method bypasses these local peculiar velocity limitations by reaching out into the deep Hubble flow. Compared to water megamasers - which are geometrically robust but relatively rare and limited in distance - this method can measure larger distances across a wider population of sources. Furthermore, while standard candles like those observed by JWST only reliably extend out to roughly 70 Mpc, our radio-geometric approach can measure distances far beyond this limit.

One of the most significant challenges in measuring cosmological distances with the SSG method is the requirement to measure the Doppler factor ($\delta$) in a cosmology-independent way. In \citet{hodgson2023}, (Paper II), it was proposed that a maximum intrinsic brightness temperature ($T_{\rm B,int}$) that is consistent across all sources could be used to correct for relativistic effects, an assumption shown to be correct within errors by \citet{cheong2025}. 

Future observations with the Square Kilometre Array (SKA) VLBI network and next generation Very Large Array (ngVLA) will heavily complement and enhance this technique. The unprecedented sensitivity of SKA-VLBI will allow us to resolve the jet and counter-jet structures in much fainter, higher-redshift AGN, exponentially increasing the sample size of viable SSG-CJ targets. Other promising independent cosmological distance methods include gravitational wave standard sirens \citep{abbott2017} and strong gravitational lensing time delays \citep{suyu2017}, which, similar to the SSG method, aim to completely bypass the traditional distance ladder calibration. Our proposed method is complementary to these methods.

In this work we propose the first direct application of a method combining Doppler factors and proper motions to determine cosmological distances, first proposed in galactic microquasars \citep{mirabel1994} and for extragalactic sources by \citet{homan2000}. An ideal source to test this method is 3C 84, as it is low redshift, has a wealth of archival data and other distance indicators to the host (Perseus) cluster. We apply this method to 3C 84 by extending the work previously published in Paper I.

\section{Methods}

In relatively rare cases where a counter-jet is detected the kinematic asymmetry ($R_{\rm kin}$) between the approaching and receding jets can be measured. Under the assumption of intrinsically identical, bidirectional jets that were launched at the same time, the apparent length ratio ($R_{\rm L}$), size ratio ($R_{\rm S}$) and/or the proper motion ratio ($R_{\rm \mu}$) can be measured (with a derivation given in Appendix \ref{app:derivation}). This allows the quantity $\beta\cos\theta_{\rm v}$ (where $\beta$ is the intrinsic speed of the jet in units of the speed of light and $\theta_{\rm v}$ is the viewing angle to the source) to be measured (e.g. \citep{fujita17}). This can then be related to the apparent approaching jet speed and the Doppler factor \citep{ghisellini93,urry95}. A fundamental assumption is that the pattern flow (observed in proper-motion measurements) and the fluid flow (which determines the Doppler factor) are equivalent \citep{homan2000}. This equality arises because the transverse kinematic information captured by the apparent jet motions $\beta_{\rm app}$ and the radial time-compression effects included in $\delta$ geometrically combine to reconstruct the full asymmetry:
\begin{equation}\label{eq:ratio_beta_dop}
    R_{\rm L} = \frac{1+\beta\cos\theta_{\rm v}}{1-\beta\cos\theta_{\rm v}}=\beta_{\rm app}^{2}+\delta^{2},
\end{equation}
and $\beta_{\rm app}$ is the apparent speed of the approaching jet:
\begin{equation}\label{eq:beta_app}
    \beta_{\rm app} = \frac{\mu D_{\rm A} (1+z)}{c},
\end{equation}
where $\mu$ is the proper motion (e.g. mas/year) of the jet. Here, we use the apparent length ratio, $R_{\rm L}$. Ideally, $R_{\mu}$ could be also simultaneously measured and should be consistent with $R_{\rm L}$ if our underlying assumptions are accurate. $R_{\rm S}$ can also be used with some additional assumptions (see Appendix \ref{app:derivation} for details). An alternative is to use the flux ratio, which is more flexible but requires further assumptions about the discrete or continuous nature of jet components \citep{homan2000}. The preference for using the size ratio over the flux ratio in this case is discussed by \citet{fujita17} . By rearranging Eq. \ref{eq:DA_original} for $\delta$ and combining with Eq. \ref{eq:beta_app} we can measure the angular diameter distance:
\begin{equation}
    D_{\rm A} = \sqrt{\frac{R_{\rm L} }{\mu^{2}+(g^{2}\theta_{\rm VLBI}^{2}/\kappa^{2}\Delta t^{2})}}\frac{c}{(1+z)}.
\end{equation}
We dub this method for measuring distances the `standard speed-gun counter-jet' (SSG-CJ). A schematic of how the method works is given in Fig. \ref{fig:method_geometry}.
\section{Results}
\label{sec:sec}
\subsection{3C 84}
\begin{figure}[htbp]
    \centering
    \includegraphics[width=1.0\linewidth]{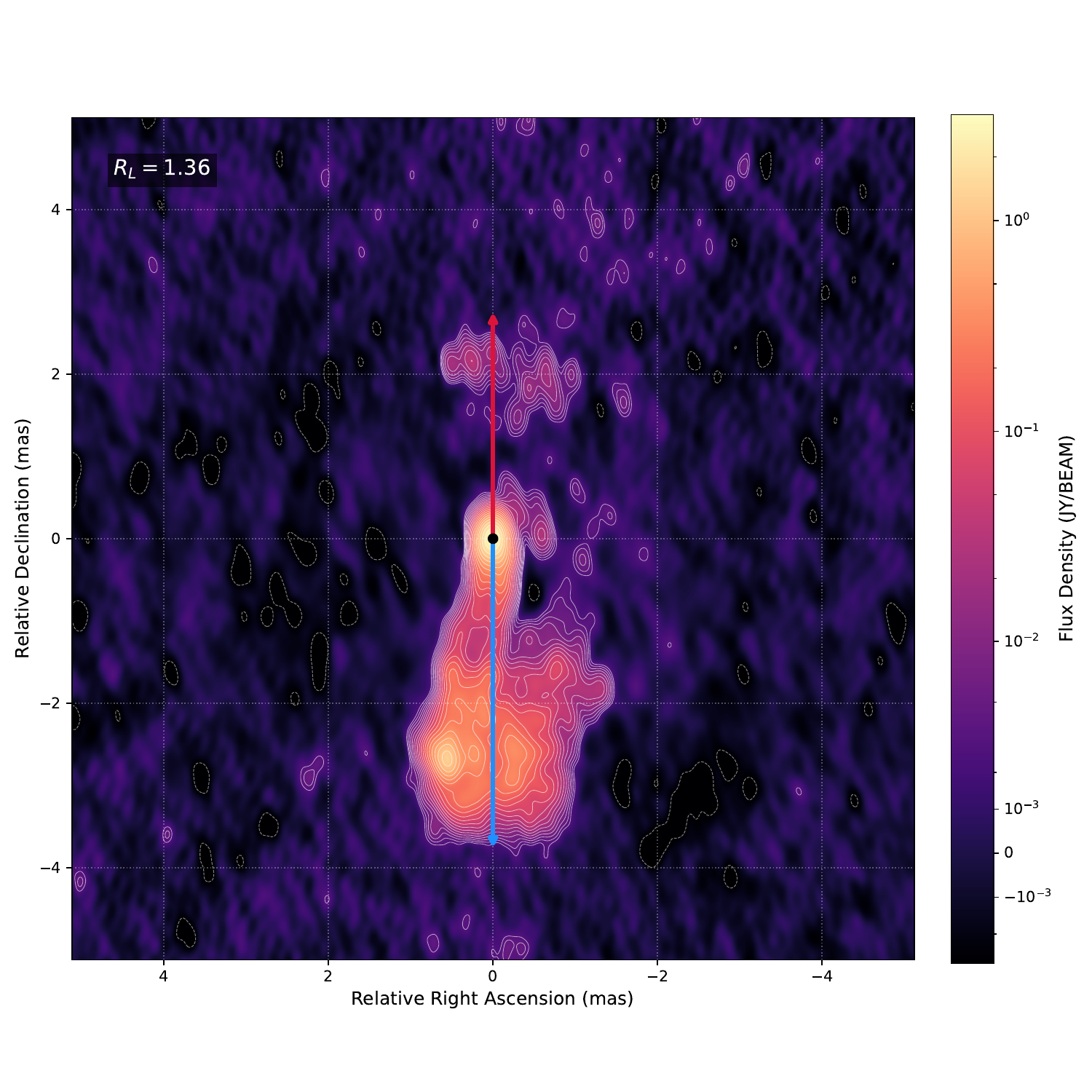}
\caption{ \textbf{43 GHz VLBI radio map of 3C 84 demonstrating the kinematic overlay of the Standard Speed-Gun Counter-Jet (SSG-CJ) method.} The image axes are given in relative milliarcseconds (mas) with the assumed AGN core centered at (0,0), marked by the black circle. The background colormap displays the flux density, with overlaid contours starting at $3\sigma$ and increasing by factors of $\sqrt{2}$. The overlaid vectors illustrate the apparent length of the approaching jet (blue, extending South) and the receding counter-jet (red, extending North). By measuring the extent of this asymmetry, we calculate an apparent length ratio of $R_L = 1.36$.}    \label{fig:method_3c84}
\end{figure}

We use the measured values of $\Delta t$=146$\pm$5 days and $\theta_{\rm VLBI}$=0.2$\pm$0.02 mas from Paper I and $\mu$=0.44$\pm$0.1 mas/yr from \citet{hodgson2021,kam2024}. \citet{fujita17} reported $R_{\rm L}$ of 1.22$\pm$0.16 using a 43 GHz VLBI image that was taken during the relevant flare; however, they assumed that the jet is being launched from the brightest point in the 43 GHz jet and assumed any core-shift effect to be negligible. Assuming a conical jet profile in equipartition, \citet{paraschos2021} reported that the jet apex is likely 0.122$\pm$0.072 mas upstream of the brightest component in the 43 GHz image. Incorporating this shift, we recalculated the ratio as $R_{\rm L}$=1.36$\pm$0.25. A preliminary check of the counter-jet proper motion yielded a ratio consistent with the measured length ratio, although the detection is not sufficiently robust for quantitative use. An image of the source with the length ratio denoted is shown in Fig. \ref{fig:method_3c84}.

\begin{table}[ht!]
    \caption{Comparison of physical assumptions and derived distances}
    \label{tab:comparison}
    \centering
    \begin{tabular}{lccc}
        \hline\hline
        Physical Assumption & $\kappa$ & Geometry ($g$) & $D_{A}$ (Mpc)  \\
        \hline
        Diameter       & 0.50 & 1.6 (Disk)   & $\approx 39.5$ \\
        Radius        & 1.00 & 1.8 (Sphere) & $\approx 71.2$   \\
        Radius        & 1.00 & 1.6 (Disk)   & $\approx 78.9$   \\
        Slow Prop. (Radius) & 1.10 & 1.8 (Sphere) & $\approx 78.3$  \\
        Slow Prop. (Radius) & 1.10 & 1.6 (Disk)   & $\approx 86.7$  \\
        Elongated (Radius) & 0.9  & 1.8 (Sphere) & $\approx 64.1$ \\
        Elongated (Radius) & 0.9  & 1.6 (Disk)  & $\approx 71.0$  \\
        \hline
    \end{tabular}
\end{table}
Errors were propagated using standard Monte Carlo techniques, simulating each variable with a normal distribution. Due to the non-linear dependence of $D_{\rm A}$ on the proper motion $\mu$ in the denominator, the resulting posterior distribution for the distance exhibits a positive skew. We report the median of the Monte Carlo distribution and the 16th and 84th percentiles as our confidence intervals. The derived distance is highly sensitive to the assumed geometry ($g$) and the causality correction factor ($\kappa$), which represents the ratio between the source radius and the light-travel time distance. As detailed in Table \ref{tab:comparison}, we explored how variations in these parameters affect the measured distance and derived cosmology.
Assuming $\kappa=1$ (radius), we produced two distinct solutions: $g=1.6$ (disk) and $g=1.8$ (sphere). This yields angular diameter distances to 3C 84 of $D_{\rm A, disk} = 78.9_{-9.8}^{+11.0}$ Mpc or $D_{\rm A, sphere} = 71.2_{-8.8}^{+9.7}$ Mpc at the reported redshift of $z=0.0178$. These distances correspond to estimated Hubble Constants of $H_{0,\rm disk}=68.5^{+6.9}_{-10.0}$ km s$^{-1}$ Mpc$^{-1}$ or $H_{0,\rm sphere}=75.8^{+8.1}_{-10.6}$ km s$^{-1}$ Mpc$^{-1}$, respectively.
Table \ref{tab:comparison} further illustrates that if the variability timescale corresponds to a diameter rather than a radius ($\kappa=0.5$), the derived distance is approximately halved, creating a significant tension with all literature measurements. Thus, we treat the radius as a lower limit, implying $\kappa \geq 1$ in a spherically symmetrical case. Conversely, if the signal propagation is sub-luminal ($\kappa=1.1$) or the source is elongated ($\kappa=0.9$), the distances shift accordingly, highlighting the need to carefully constrain these astrophysical factors.
\subsection{Comparison with literature distances}
3C 84 is the central radio galaxy of the Perseus cluster and also hosts Type Ia supernovae (SN Ia), which could both offer useful independent distances. SN 2005mz belongs to the 1991bg-like subclass, which are intrinsically subluminous and evolve faster than normal Type Ia SNe. Modern standardization frameworks, such as those used in the Pantheon+ sample \citep{scolnic2022pantheon}, explicitly exclude these objects due to their deviation from the Phillips relation \citep{phillips1993absolute}. While \citet{hicken2009} applied the SALT model to derive a distance modulus, SALT was not trained on the unique spectral time-series of 91bg-like events. Consequently, the derived distance of $\sim 74$ Mpc carries unquantifiable systematic uncertainties beyond the statistical errors. Therefore, we use this distance solely as a consistency check rather than a rigid calibrator.
As a comparison, we can also consider the distance to the Perseus cluster itself as a proxy, especially since 3C 84 is the radio core of NGC 1275, which is the brightest galaxy in the cluster.  Therefore, this distance should be representative of the distance to the galaxy. The `distance' to the Perseus cluster was measured using the Fundamental Plane relation by \citet{hudson97} to be $5176\pm185$ km/s. This distance was measured relative to the Coma cluster, whose distance was set to $cz_{\text{CMB}}=7200$ km/s (its `redshift distance' with no peculiar velocity). Assuming the SH0ES $H_0=73.04\pm1.04$ km/s/Mpc, consistent with the case of SN 2005mz, the distance is $D_A=69.6\pm2.7$ Mpc. 
Conversely, if we assume a lower value of $H_0$ (e.g., $\sim 67.4$ km s$^{-1}$ Mpc$^{-1}$), the physical distance to the cluster would \textit{increase} (as $D \propto v/H_0$) to approximately $75.4$ Mpc. This illustrates the degeneracy: lower values of $H_0$ imply larger physical distances for a given redshift/velocity.

\subsubsection*{\textbf{Implications assuming the SN Ia (SH0ES) Distance}}
If we assume the SN Ia distance ($D_A \approx 74$ Mpc) is correct, we can actively constrain the jet physics. For the optically thin component in 3C 84, this distance favors a spherical geometry with $\kappa \approx 1$. A disk-like geometry would require $\kappa < 1$, implying the source size is closer to a diameter than a radius, which is physically less likely. Furthermore, this actively rules out sound-speed-limited expansion models for this specific flare in 3C 84, which would require $\kappa \approx 0.57$ and yield a highly discrepant distance of roughly 40--45 Mpc.

\subsubsection*{\textbf{Implications assuming a lower $H_0$ Calibration}}
If a lower $H_0$ calibration is correct, the expected distance is $\sim$77.5 Mpc. This scenario is highly compatible with either a disk-like geometry with $\kappa \sim 1$ or a spherical geometry with a slight causality correction of $\kappa \sim 1.1$ (e.g., due to mild sub-luminal expansion or elongation).
\section{Discussion}
\label{sec:disc}
The results highlight the potential of the SSG-CJ method, but also the dependence on geometric assumptions. The consistency of our results with external indicators is highly dependent on the choices of $g$ and $\kappa$.
\subsection{Systematics of Free-Free Absorption}
We must note a crucial systematic effect regarding the host environment of NGC 1275: free-free absorption. The northern counter-jet in 3C 84 is viewed through an accretion disk or torus, which causes severe free-free absorption at cm-wavelengths \citep{walker2000}. This issue was a motivation for \citet{fujita17} to use the length ratio over the flux density ratio, as the observed flux density ratio was much higher ($\sim$45 at 43 GHz) than the expected $\sim$1.8-2.1 due to free-free absorption. 

While observing at 43 GHz minimizes opacity effects compared to lower frequencies, if the counter-jet is partially absorbed, its apparent length is artificially truncated. This means our measured $R_{\rm L}$ of 1.36 might be a lower limit. A larger intrinsic $R_{\rm L}$ would systematically increase the derived distance $D_{\rm A}$, pushing the result further away from the SH0ES $H_0$ calibration and closer to the CMB $H_0$ calibration. Higher frequency observations, such as at 86 GHz with the Global Millimeter VLBI Array (GMVA) \citep{oh2022,paraschos2021,paraschos2024}, would definitively resolve this opacity issue. However, since ideally all measurements would be made at the same frequency, higher cadence observations would be required than the $\sim$6 monthly schedule of the GMVA.

\subsection{Distance anchor or Astrophysical Probe?}

The utility of the SSG-CJ method depends on which variables can be constrained. We consider two limiting cases:

\begin{enumerate}
    \item In the best-case scenario, if we can solve for the geometric factors $g$ and $\kappa$ solely from internal data (or theoretical priors), the SSG-CJ method becomes a primary distance indicator - a "single rung" distance measure that does not require calibration via Cepheids or TRGB. Our result of $D_{\rm A} \approx 71.2$ Mpc (sphere) is consistent with SH0ES-calibrated indicators, suggesting that if $\kappa \approx 1$, this method could independently support a higher local $H_0$.
    
    \item In the worst-case scenario, where geometry cannot be internally determined, the method is inverted. By adopting precise external distance measurements (e.g., from future high-precision SN Ia or gravitational wave standard sirens), we can use the SSG-CJ framework to investigate the distribution of jet geometries ($g$) and causality factors ($\kappa$). This would provide unique constraints on the physics of jet launching and particle acceleration.
\end{enumerate}

\subsection{Consistency with literature distance estimates}

Both the SN Ia and FP distances derived above ultimately rely on the higher calibrated value of $H_0$ and are consistent with each other. These distances are broadly consistent with our SSG-CJ distance of $D_{\rm A} = 71.2_{-8.8}^{+9.7}$ Mpc, assuming a spherical geometry ($g=1.8$) and $\kappa=1$. This value aligns with a model angular diameter distance of $D_{\rm A,model,high} = 71.5$ Mpc (assuming $H_0 \approx 73.0$ km s$^{-1}$ Mpc$^{-1}$).

On the other hand, if a lower $H_0$ cosmology is correct ($H_0 \approx 67.4$ km s$^{-1}$ Mpc$^{-1}$), the expected distance to 3C 84 is $D_{\rm A,model,low} \approx 77.5$ Mpc. Our data can accommodate this larger distance if we assume a disk-like geometry ($D_{\rm A, disk} \approx 78.9$ Mpc). Thus, the distinction between supporting a higher or lower $H_0$ cosmology currently rests on distinguishing between a spherical or disk-like geometry for the emitting region. It should also be noted that a larger sample could also mitigate the degeneracy. 

\section{Conclusions}

We have presented the SSG-CJ method, a refinement of the standard speed-gun distance that eliminates the dependency on an assumed Doppler factor by utilizing counter-jets and kinematics. Applied to 3C 84, we derive distances consistent with literature values, though the precision is currently limited by geometric assumptions.

As in Paper I, these results serve primarily as a proof-of-concept. A clear limitation is the assumption of identical approaching and receding jets. While unrealistic for individual epochs, this assumption should hold statistically over large samples.

Currently, there is a relative lack of sources that exhibit variability, super-luminal motions and counter-jets. Measuring distances in this way represents an ideal synergy between high-cadence, single-dish monitoring programs (to accurately determine $\Delta t$) and ultra-high sensitivity snapshot arrays like SKA-VLBI and ngVLA (to resolve the jet/counter-jet kinematics $R_{\rm L}$ and $\mu$). Long-wavelength VLBI, where counter-jets are intrinsically brighter and less obscured by inner-disk opacity, will be critical.

A limitation of using a maximum intrinsic brightness temperature to calibrate relativistic effects is that it assumes that the value is consistent across all sources and does not evolve with redshift, assuming the brightness temperature limit is frequency-independent in the rest frame. As most counter-jet sources are likely to be at low redshift, it could be possible to measure the intrinsic brightness temperature at multiple frequencies and apply the values at high-redshift - e.g. a 43 GHz $T_{\rm B,int}$ measurement at low-$z$ should be equivalent to 22 GHz measurement at $z=1$ or a 15 GHz measurement at $z=2$.

In the future, we plan to extend this analysis to other low-redshift sources and at multiple frequencies. If the uncertainties regarding the geometry can be effectively resolved, then the SSG-CJ method will be able to concretely contribute to the debate regarding the Hubble tension and the potential hints of a time variable Dark Energy.

\begin{acknowledgements}
J.A.H. acknowledges the support of the National Research Foundation of Korea (NRF) (NRF-2021R1C1C1009973) and that this work was supported by the National Research Foundation of Korea (NRF) grant funded by the Korea government(MSIT) RS-2025-16302968. SHOH acknowledges that  this work was supported by the National Research Foundation of Korea (NRF) grant funded by the Korea government(MSIT) (RS-2026-25469119). SHOH acknowledges support from the National Research Foundation of Korea (NRF) grant funded by the Korea government (Ministry of Science and ICT: MSIT) (No. RS-2026-25469119). IL was funded by the European Union ERC-2022-STG - BOOTES - 101076343. Views and opinions expressed are however those of the author(s) only and do not necessarily reflect those of the European Union or the European Research Council Executive Agency. Neither the European Union nor the granting authority can be held responsible for them. This research was partially funded by the Deutsche Forschungsgemeinschaft (DFG, German Research Foundation) as part of the DFG Research Unit FOR5195 – project number 443220636.

\end{acknowledgements}

\bibliographystyle{aa}
\bibliography{oja_template}

\begin{appendix}
\section{Derivation of Kinematic Asymmetry Ratios}
\label{app:derivation}

While the flux ratio between an approaching and receding jet is a standard metric in the literature \citep[e.g.,][]{homan2000, ghisellini93, urry95}, spatial and kinematic ratios—such as the length ratio ($R_{\rm L}$), proper motion ratio ($R_{\mu}$), and size ratio ($R_{\rm S}$)—provide direct geometric constraints when the approaching and receding components are assumed to be intrinsically identical and ejected at the same time. Here we demonstrate that these observable ratios share an identical dependence on the apparent velocity ($\beta_{\rm app}$) and the Doppler factor ($\delta$), such that $R_{\rm L} = R_{\mu} = R_{\rm S} = \beta_{\rm app}^2 + \delta^2$.

Consider a symmetric twin-jet system where identical components are ejected simultaneously with an intrinsic velocity $\beta c$ at a viewing angle $\theta_{\rm v}$ to the observer's line of sight.

\subsection*{Derivation of the Kinematic Asymmetry Ratio}
Over a given intrinsic source-frame time $t$, the physical distance travelled by each component in the host galaxy frame is $L = \beta c t$. Due to light-travel time effects, the apparent length of the approaching jet ($l_{\rm app}$) and the receding counter-jet ($l_{\rm rec}$) projected on the plane of the sky are:
\begin{equation}
    l_{\rm app} = \frac{L \sin\theta_{\rm v}}{1 - \beta \cos\theta_{\rm v}}, \quad \text{and} \quad l_{\rm rec} = \frac{L \sin\theta_{\rm v}}{1 + \beta \cos\theta_{\rm v}}.
\end{equation}
The apparent length ratio is therefore:
\begin{equation}
    R_{\rm L} = \frac{l_{\rm app}}{l_{\rm rec}} = \frac{1 + \beta \cos\theta_{\rm v}}{1 - \beta \cos\theta_{\rm v}}.
    \label{eq:app_length_ratio}
\end{equation}

Because proper motion is simply the apparent transverse velocity divided by angular diameter distance ($\mu \propto v_{\rm app} = l_{\rm app} / \Delta t_{\rm obs}$), it is governed by identical photon arrival-time delays, yielding $R_{\mu} = \mu_{\rm app}/\mu_{\rm rec} = R_{\rm L}$. Similarly, for structures whose characteristic physical size scales linearly with time or distance from the apex (e.g., self-similarly expanding knots or conical jets), the observed angular size scales directly with apparent length, making the size ratio $R_{\rm S} = \theta_{\rm app}/\theta_{\rm rec}$ identical to $R_{\rm L}$ under these model-dependent geometric assumptions.

\subsection*{Conversion to Observable Parameters}
Given these equivalences, we define a generalized observable kinematic ratio $R_{\rm kin} = (1 + \beta \cos\theta_{\rm v}) / (1 - \beta \cos\theta_{\rm v})$. We can express this ratio entirely in terms of the apparent transverse velocity ($\beta_{\rm app}$) and the approaching Doppler factor ($\delta$), defined as:
\begin{equation}
    \beta_{\rm app} = \frac{\beta \sin\theta_{\rm v}}{1 - \beta \cos\theta_{\rm v}}, \quad \text{and} \quad \delta = \frac{\sqrt{1 - \beta^2}}{1 - \beta \cos\theta_{\rm v}}.
\end{equation}

Summing the squares of these two quantities yields:
\begin{equation}
    \beta_{\rm app}^2 + \delta^2 = \frac{1 + \beta \cos\theta_{\rm v}}{1 - \beta \cos\theta_{\rm v}}.
\end{equation}

This derivation demonstrates that kinematic asymmetries directly constrain the Doppler factor when the apparent velocity is known, yielding the relationship central to the SSG-CJ method:
\begin{equation}
    R_{\rm kin} = R_{\rm L} = R_{\mu} = R_{\rm S} = \beta_{\rm app}^2 + \delta^2,
\end{equation}
where $R_{\rm S}$ requires the additional assumption of linear physical expansion.

\end{appendix}
\end{document}